\documentclass[aps,prd,nofootinbib,showpacs,floatfix,superscriptaddress,preprintnumbers]{revtex4}
\usepackage{amsmath}
\usepackage{amssymb}
\usepackage{epsfig}
\usepackage{graphicx}
\usepackage{pstricks}
\DeclareMathAlphabet{\mathsc}{OT1}{cmr}{m}{sc}
\newcommand {\ignore}[1]{}

\def\10{$SO(10)$}
\def\21{SU(2) $\otimes$ U(1) }

\def\422{$SU(4) \otimes SU(2) \otimes SU(2)$}
\def\321{SU(3) $\otimes$ SU(2) $\otimes$ U(1)}
\def\gsim{\raise0.3ex\hbox{$\;>$\kern-0.75em\raise-1.1ex\hbox{$\sim\;$}}}
\def\lsim{\raise0.3ex\hbox{$\;<$\kern-0.75em\raise-1.1ex\hbox{$\sim\;$}}}

\def\lsim{\raise0.3ex\hbox{$\;<$\kern-0.75em\raise-1.1ex\hbox{$\sim\;$}}}
\def\gsim{\raise0.3ex\hbox{$\;>$\kern-0.75em\raise-1.1ex\hbox{$\sim\;$}}}
\def\vev#1{\left\langle #1\right\rangle}
\def \znbb {0\nu\beta\beta}

\newcommand{\AddrAHEP}{%
  AHEP Group, Institut de F\'{\i}sica Corpuscular --
  C.S.I.C./Universitat de Val{\`e}ncia \\
  Edificio Institutos de Paterna, Apt 22085, E--46071 Valencia, Spain}

\newcommand{\Adderoma}{%
  Dipartimento di Fisica "E. Amaldi", \\Universit\'a degli Studi Roma Tre, Via della Vasca Navale 84, 00146 Roma }

 \newcommand{\ba}{\begin{array}}
\newcommand{\ea}{\end{array}}
\relax

\def\321{$SU(3)\times SU(2)\times U(1)$}

\begin{document}
\preprint{IFIC/11-13}
\preprint{RM3-TH/11-2}
\renewcommand{\Huge}{\Large}
\renewcommand{\LARGE}{\Large}
\renewcommand{\Large}{\large}
\def \znbb {$0\nu\beta\beta$ }
\def \nbb {$\beta\beta_{0\nu}$ }
%\title{Dark matter from non abelian discrete group}
\title{Stability of dark matter from the $D_4\times Z_2^{f}$ flavor group}
\author{D.~Meloni} \email{meloni@fis.uniroma3.it} \affiliation{\Adderoma}
\author{S.~Morisi} \email{morisi@ific.uv.es}\affiliation{\AddrAHEP}
\author{E.~Peinado} \email{epeinado@ific.uv.es}\affiliation{\AddrAHEP}
% \affiliation{\Adderoma}

\date{\today}

\begin{abstract}
We study a model based on the dihedral group $D_4$ in which the dark matter is stabilized by the interplay between a remnant $Z_2$ symmetry, 
of the same spontaneously broken non-abelian group, and an auxiliary $Z_2^{f}$ introduced to eliminate unwanted couplings in the 
scalar potential. In the lepton sector the model is compatible with normal hierarchy only and predicts a vanishing reactor mixing angle, $\theta_{13}=0$. 
Since $m_{\nu_1}=0$, we also have a simple prediction for the effective mass in terms of the solar angle: 
$|m_{\beta\beta}|=|m_{\nu_2}| \,\sin^2\theta_{\odot}\sim 10^{-3}$ eV.
There also exists a  large portion of the model parameter space where the upper bounds on lepton flavor violating processes are not violated.
We incorporate quarks in the same scheme finding that a description of the CKM mixing matrix is possible  and that semileptonic $K$ and 
$D$ decays mediated by flavor changing neutral currents are under control.

\end{abstract}


\maketitle

\section{Introduction}
\label{intro}
We have strong evidence about the existence of dark matter (DM) \cite{Bertone:2004pz,Bertonebook}. A good 
DM candidate must be neutral and stable or with a decay length bigger
than the age of the universe and give the correct relic abundance~\cite{Taoso:2007qk}. 
There are several extensions of the standard model predicting good DM candidates; however, it turns out that 
in many models the stability of the DM is obtained introducing ad-hoc assumptions, see for example the review \cite{Hambye:2010zb}. 
Any of these models may be correct but certainly it would be desirable to provide a fundamental  explanation of the origin of the stability. 
In \cite{Hirsch:2010ru} it has been pointed out that the stability can be guaranteed by a residual $Z_2$ symmetry arising from the spontaneous breaking of a
non-abelian flavor symmetry; the same $Z_2$ also acts in the neutrino sector and has a strong impact on the phenomenology of neutrino masses and mixing.
In that model the flavor symmetry
% The stabilization mechanism can give constraints on the structure of the model and therefore 
% on the associated phenomenology. 
% In particular there could be a relationship between the neutrino phenomenology and dark matter, as first
% proposed in  \cite{Hirsch:2010ru}, where
% the stability is guarantee by a residual $Z_2$ symmetry arising from the spontaneous breaking of a
% flavor symmetry. In \cite{Hirsch:2010ru} such a 
is the group of the even 
permutations of four objects $A_4$ whose irreducible representations are three singlets and one triplet.
To avoid a direct couplings to quarks and charged leptons, the DM candidate is assigned to a triplet representation, while leptons and quarks to singlets of $A_4$.
After electroweak symmetry breaking, $A_4$ is broken into its subgroup $Z_2$ under which two component
of the triplet DM are automatically charged;
%, without assuming ad-hoc parity; 
eventually, this prevents dangerous couplings with the Higgs fields of the model. 
Such an idea has been then further studied and extended in refs.\,\cite{Meloni:2010sk,Boucenna:2011tj}.

The interplay between decaying dark matter and non-abelian discrete flavor symmetries has been considered in a number of 
subsequent papers; for instance, in \cite{Haba:2010ag,Kajiyama:2010sb,Daikoku:2010ew} non-abelian discrete 
symmetries prohibit operators that may induce too fast dark matter decay; 
in \cite{Adulpravitchai:2011ei} a non-abelian discrete symmetry (not a flavor symmetry) has been used to stabilize
the scalar DM candidate (similar to what has been discussed in the inert scalar models 
\cite{LopezHonorez:2006gr}) and the matter sector has not been considered.
%in \cite{Adulpravitchai:2011ei} only the new scalar fields of the model (not the matter fields) have been assigned to 
%irreducible representation of the group $D_3$ and the DM stability does not follow from any conserved subgroups, 
%being a consequence of the reduced number of couplings in the Lagrangian (similar to what has been discussed in the inert scalar models \cite{LopezHonorez:2006gr}).
%
% where a non-abelian discrete symmetry (not a flavor symmetry),
% is imposed by hand in order to stabilize the dark matter similarly to the inert scalar models \cite{LopezHonorez:2006gr}.
Therefore the models in   \cite{Haba:2010ag,Kajiyama:2010sb,Daikoku:2010ew} and \cite{Adulpravitchai:2011ei}
are substantially different to the idea introduced in \cite{Hirsch:2010ru}. %In the former case the

In this paper we adopt the point of view elucidated in \cite{Hirsch:2010ru,Meloni:2010sk,Boucenna:2011tj}, studying 
a flavor model where the stability of the DM is caused by the interplay between a remnant $Z_2$ symmetry of the $D_4$  group and 
an auxiliary $Z_2^{f}$ which allows to eliminate dangerous couplings in the scalar potential.
Since $D_4$ contains only singlets and doublets it is highly non-trivial
to still be able to generate the mechanism for dark matter stabilization; in addition, 
the same non-abelian symmetry also acts on leptons and quarks, giving acceptable phenomenology in both sectors.  

The relevant differences of our model compared to \cite{Hirsch:2010ru,Meloni:2010sk,Boucenna:2011tj} 
can be summarized as follows:
\begin{itemize}
\item  for the first time, we extend such a mechanism to incorporate quarks transforming under non-trivial 
representation of $D_4$; in this framework, we are able to reproduce the correct order of magnitude of the quark mixing angles, a quite remarkable result; 
\item charged leptons are non-diagonal (with hierarchy among the eigenvalues naturally reproduced with 
${\cal O}(1)$ Yukawa couplings) and completely responsible for the atmospheric mixing angle in the neutrino sector, instead of being diagonal as in 
\cite{Hirsch:2010ru,Meloni:2010sk,Boucenna:2011tj};
\item although the Higgs sector is extended with three more scalar doublets and one singlet, 
the neutrino sector contains only two right-handed neutrinos. The model can be considered minimal in this respect. 
\end{itemize}

The paper is organized as follows: in section \ref{model} we present the relevant features of the model, 
discussing the group properties of $D_4$ and the assignments of leptons and Higgs fields to the irreducible representations
of the group.  In section \ref{scalar-stability} we discuss the scalar potential of the theory and describe in details 
how the DM stability arises in our model;  sections \ref{chargedsec} and \ref{fcnc} are devoted to the neutrino phenomenology and to
the estimate of some relevant lepton flavor violating processes, respectively. In section \ref{quarks} we discuss the 
quark sector and give an order of magnitude estimate 
of some of 
the flavor changing neutral current processes; eventually, in section \ref{conclusions} we 
draw our conclusions.

\section{Model}
\label{model}

We assign the fields of the model into irreducible representation of $D_4$, the dihedral group of order four \cite{Blum:2007nt},
see also \cite{Ishimori:2010au}. It has five irreducible representations, four singlets $1_{1,2,3,4}$ and one doublet $2$. 
The generators of the group fulfill the relations:
\begin{equation}
\begin{array}{cc}
A^4=B^2=1,&ABA=B.
\end{array}
\end{equation}

The one-dimensional representations are characterized by $A=B=1$ for $1_1$, $A=1$, $B=-1$ for $1_2$, $A=-1$, $B=1$ for $1_3$ and $A=-1$, $B=-1$ for $1_4$. 
The generators for the two-dimensional representations are
\begin{equation}\label{generators}
\begin{array}{cc}
A=\left(\begin{array}{cc}i&0\\0&-i\end{array}\right),&B=\left(\begin{array}{cc}0&1\\1&0\end{array}\right).
\end{array}
\end{equation}
An interesting feature of $D_4$ is that the product of two doublets contains only singlets:
given $(a_1,a_2)\sim 2$ and $(b_1,b_2)\sim 2$ we have:
\begin{equation}
\begin{array}{ll}
a_1b_2+a_2b_1\sim 1_1, &\quad a_1b_2-a_2b_1\sim 1_2,\\
a_1b_1+a_2b_2\sim 1_3, &\quad a_1b_1-a_2b_2\sim 1_4.
\end{array}
\end{equation}
For the singlets: 
$1_i\times1_i=1_1,~1_1\times1_i=1_i$ for $i=1,\cdots,4$, $1_2\times 1_3 = 1_4$,
$1_2\times 1_4 = 1_3$ and $1_3\times 1_4 = 1_2$. 
The standard model Higgs doublet is taken as a singlet $1_1$; we 
assume three further Higgs doublets, one of them transforming as a singlet $1_3$ ($H'$) and the other as a doublet of $D_4$, $\eta=(\eta_1,\eta_2)\sim 2$.  
In order to correctly describe both lepton and quark sectors, we need to introduce a scalar $SU(2)$  singlet flavon $\phi$ in the $1_2$ representation. 
Two right-handed neutrinos ($N_1$, $N_2$) in the doublet representation $N_D$ are necessary ingredients to give mass to the neutrinos via the type-I seesaw 
mechanism. The assigment to the irreducible representations of $D_4$ as well as the charges under $Z_2^{f}$ are listed in 
Tab.\ref{tab1}.
\begin{table}[h!]
\begin{center}
\begin{tabular}{|c|c|c|c|c|c|c|c||c|c|c|c|}
\hline
&$L_e$&$L_{\mu}$&$L_{\tau}$&$l_{e}^c$&$l_{{\mu}}^c$&$l_{{\tau}}^c$&$N_{D}$&$H$&$H'$&$\eta$&$\phi$\\
\hline
$SU(2)$&2&2&2&1&1&1&1&2&2&2&1\\
\hline
$D_4$ &$1_1$ &$1_2$&$1_3$&$1_4$&$1_1$&$1_3$&$2$ &$1_1$ &$1_3$&2&$1_2$\\
\hline
$Z_2^{f}$ &$+$ &$+$&$+$&$+$&$-$&$+$&$+$&$+$ &$-$&$+$&$-$\\
\hline
\end{tabular}\caption{\it Assignment of the lepton and Higgs fields under $SU(2)$, $D_4$ and $Z_2^{f}$.}\label{tab1}
\end{center}
\end{table}

The invariant $D_4\times Z_2^f$ Lagrangian in the lepton sector is as follow:
\begin{eqnarray}
\label{lag}
\mathcal{L}&=&\frac{y_{1}^l}{\Lambda} L_el_{_e}^c H^\prime \phi+\frac{y_{2}^l}{\Lambda}  L_\mu l_{\mu}^c H \phi+y_{3}^l L_\tau l_{\tau}^c H+
\frac{y_{4}^l}{\Lambda} L_\mu l_{\tau}^c H^\prime\phi+
y_{5}^l L_\tau l_{\mu}^c H^\prime+\nonumber\\
&&y_1^\nu L_e(N_D\eta)_{1_1}+y_2^\nu L_\mu(N_D\eta)_{1_2}+y_3^\nu L_\tau(N_D\eta)_{1_3}+\\
&&+M_1 N_DN_D+
\mbox{h.c.}\nonumber\,,
\end{eqnarray}
where $\Lambda$ is a large energy scale. Here we have only considered terms up to one-flavon insertion. It turns out that 
higher powers of Higgs fields and flavon insertions can only modify (to a negligible level) the couplings $y_i^l$ but are
not able to generate new Yukawa interactions. In the neutrino sector, due to the $Z_2^{f}$ symmetry, the previous lagrangian is modified by operators with at least two
powers of $\phi$, that we assume here negligible. 
We will investigate in details its phenomenological consequences in sections \ref{chargedsec} and \ref{fcnc}.

\section{Scalar spectrum and stability of the dark matter candidate}
\label{scalar-stability} 

The invariant scalar potential is of the form:
\begin{equation}\begin{array}{lcl}
V&=&\mu_H^2 H^\dagger H+\mu_{H^\prime}^2 H^{\prime\dagger} H^\prime+\mu_{\eta}^2(\eta^\dagger \eta)_{1_1}+\mu_\phi \phi^2+\\\\
&+&\lambda_1 (\eta^\dagger \eta)_{1_1} (\eta^\dagger \eta)_{1_1}+\lambda_2 (\eta^\dagger \eta)_{1_2} (\eta^\dagger \eta)_{1_2}+
\lambda_3 (\eta^\dagger \eta)_{1_3} (\eta^\dagger \eta)_{1_3}+\lambda_4 (\eta^\dagger \eta)_{1_4} (\eta^\dagger \eta)_{1_4}+\\\\
&+&\lambda^\prime_1 (\eta^\dagger \eta^\dagger)_{1_1} (\eta  \eta)_{1_1}+\lambda^\prime_2 (\eta^\dagger \eta^\dagger)_{1_2} (\eta  \eta)_{1_2}+\lambda^\prime_3 (\eta^\dagger \eta^\dagger)_{1_3} (\eta  \eta)_{1_3}+\lambda^\prime_4 (\eta^\dagger \eta^\dagger)_{1_4} (\eta  \eta)_{1_4}+\lambda_5 (\eta^\dagger \eta)_{1_1}H^\dagger H+\\\\
&+&\label{pote}\lambda_5^\prime (\eta^\dagger H)(H^\dagger \eta)+\lambda_6 [(\eta^\dagger \eta^\dagger)_{1_1}H H+\mbox{h.c.}]+
\lambda_7 (\eta^\dagger \eta)_{1_1}H^{\prime\dagger} H^\prime+\lambda_7^\prime (\eta^\dagger H^\prime)(H^{\prime\dagger} \eta)+
%\lambda_7[(\eta^\dagger \eta)_{1_3}H^\dagger H^\prime+\mbox{h.c.}]+
%\lambda_7^\prime[(\eta^\dagger H^\prime )(H^\dagger\eta) +\mbox{h.c.}]+
\\\\&+&
%\lambda_8[(\eta^\dagger \eta^\dagger)_{1_3}(HH^\prime) +\mbox{h.c.}]+
\lambda_{8}[(\eta^\dagger \eta^\dagger)_{1_1}H^\prime H^\prime+\mbox{h.c.}]+
\lambda_{9}(H^{\prime \dagger}H^\prime)(H^\dagger H)+\lambda_{9}^\prime(H^{\prime \dagger}H)(H^\dagger H^\prime)+
\\\\&+&\lambda_{10}[(H^{\prime \dagger}H)(H^{\prime \dagger}H)+\mbox{h.c.}]+\lambda_{11}(H^\dagger H)(H^\dagger H)+
\lambda_{12}(H^{\prime \dagger}H^\prime)(H^{\prime \dagger}H^\prime)+\\\\&+&\xi_1\phi^2H^\dagger H+\xi_2\phi^2H^{\prime\dagger} H^\prime+\xi_3\phi^2(\eta^\dagger \eta)_{1_1}+\xi_4 \phi^4.
\end{array}\end{equation}
We assume a vev structure of the form:
\begin{equation}
\label{vevs}
\begin{array}{ccccc}
\langle H \rangle=v,&
\langle H^\prime \rangle=v^\prime,&
\langle \eta_1\rangle =v_{\eta_1},&
\langle \eta_2\rangle=v_{\eta_2},&
\langle \phi\rangle=v_{\phi}
\end{array}
\end{equation}
where the various vevs $v_i$ are obtained solving the coupled differential equations
$\partial V/\partial v_i=0$.  Assuming for simplicity real vevs,
we have carefully checked that, for suitable parameter choices of the potential
$V$, an allowed  local minimum is:
\begin{equation}
v_{\eta_1}=v_{\eta_2}=v_{\eta},
\end{equation}
which is the crucial point to justify the stability of the DM based on symmetry arguments.
After electroweak symmetry breaking we can write:
\begin{eqnarray}
\label{etas}
\eta_1&=&\left(
\begin{array}{c}
\eta^+_1\\
v_{\eta}+\eta_1^\prime+iA_1
\end{array}
\right),\qquad
\eta_2=\left(
\begin{array}{c}
\eta^+_2\\
v_{\eta}+\eta_2^\prime+iA_2
\end{array}
\right),\\
H&=&\left(
\begin{array}{c}
H^+\\
v+H+i A
\end{array}
\right),\qquad H^\prime=\left(
\begin{array}{c}
H^{\prime +}\\
v^\prime+H^\prime+i A^\prime
\end{array}
\right),
\end{eqnarray}
and the physical spectrum involves  four neutral scalars, three pseudoscalars
and three charged scalars { (plus one flavon)}. 
To maintain the notation compact and avoid unnecessary complications, we work in the limit of decoupled $\phi$ (that is $\xi_{1,2,3}=0$),
which does not modify any of the results discussed in the paper (and, of course, the vev alignment $\langle \eta \rangle \sim (1,1)$).
In such a limit,  the mass matrices of the three sectors ($S$=scalar, $A$=pseudoscalar, $H^+$=charged) can be generically
written in the following way:
\begin{equation}
(M^{S,A,H^+})^2=\left(\begin{array}{cccc}M^{S,A,H^+}_{11}&M^{S,A,H^+}_{12}&M^{S,A,H^+}_{13}&M^{S,A,H^+}_{13}\\M^{S,A,H^+}_{12}&M^{S,A,H^+}_{22}&
M^{S,A,H^+}_{23}&M^{S,A,H^+}_{23}\\M^{S,A,H^+}_{13}&M^{S,A,H^+}_{23}&M^{S,A,H^+}_{33}&M^{S,A,H^+}_{34}\\M^{S,A,H^+}_{13}&M^{S,A,H^+}_{23}&
M^{S,A,H^+}_{34}&M^{S,A,H^+}_{33}\end{array}\right).\label{higgsmass}
\end{equation}
The relevant feature here is that the $2\times 2$ sub-block corresponding to the  $3-4$ sector is symmetric and  
can be put in a block diagonal form by a maximal rotation. This corresponds to a rotation in the corresponding bidimensional
% ($\eta_1^\prime$, $\eta_2^\prime$) 
subspace
which defines the mass eigenstates of the subsector. 
After this change of basis, we are left with block-diagonal mass matrices made by
$3\times3$ matrices (one for scalars, one for pseudoscalars and one for charged), and $1\times1$ blocks 
corresponding to the isolated DM sector:
%(DA VERIFICARE!! DEFINIRE m)
\begin{equation}
M^{S,A,H^+}=
\left(\begin{array}{cccc}
M^{S,A,H^+}_{11}&M^{S,A,H^+}_{12}&\sqrt{2}M^{S,A,H^+}_{13}&0\\
M^{S,A,H^+}_{12}&M^{S,A,H^+}_{22}&\sqrt{2}M^{S,A,H^+}_{23}&0\\
\sqrt{2}M^{S,A,H^+}_{13}&\sqrt{2}M^{S,A,H^+}_{23}&M^{S,A,H^+}_{33}+M^{S,A,H^+}_{34}&0\\
0&0&0&(m^{S,A,H^+}_{DM})^2
\end{array}\right),\label{higgsmassrot}
\end{equation}
where 
\begin{equation}
\label{dmmass}
(m^{S,A,H^+}_{DM})^2=M^{S,A,H^+}_{33}-M^{S,A,H^+}_{34}\,.
\end{equation}
% and
% \begin{equation}\begin{array}{l}
% (m^{S}_{DM})^2=-2 \left(2 (\lambda_1^\prime-\lambda_2-\lambda_3^\prime+\lambda_4) v_\eta^2+\lambda_6 v^2+ 
% %(\lambda_7+\lambda_7^\prime) v v^\prime+
% \lambda_{10} v^{\prime 2}\right)\\
% (m^{A}_{DM})^2= -4 (\lambda_3-\lambda_3^\prime+\lambda_4+\lambda_4^\prime)v_\eta^2
% %-2 (\lambda_7+\lambda_7^\prime+\lambda_8)vv^\prime
% \\
% (m^{H^+}_{DM})^2=\frac{1}{2} \left(-4 (\lambda_1^\prime+\lambda_2^\prime-\lambda_3^\prime+2 \lambda_4+\lambda_4^\prime) v_\eta^{\prime 2}-2 (\lambda_5^\prime+\lambda_6) v^2
% %-2 (2 \lambda_7+\lambda_7^\prime+\lambda_8)v v^\prime
% -(2 \lambda_{10}+\lambda_9^\prime) v^{\prime 2}\right).
% \end{array}\label{masses}\end{equation}
The explicit expressions of the physical masses are complicated functions of the potential parameters and their expressions do not reveal any
important features to be mentioned here beside the fact that, as expected, the pseudoscalars and charged mass matrices 
have a zero eigenvalues corresponding to the Goldstone bosons.

We have verified that, for a suitable set of the potential parameters, $m^S_{DM}$ is the lightest mass of the 
neutral stable states and it is smaller than the corresponding charged states carrying the same 
$Z_2$ parity; in addition, all the masses arising from the diagonalization of the $3\times 3$ matrices in 
eq.(\ref{higgsmassrot}) are positive and do not violate the electroweak constraints on the $T,S$ and $U$ 
oblique parameters\,\cite{Grimus:2008nb}.

We are now in the position to discuss the mechanism for the DM stability in our model. 
The condition in eq.(\ref{vevs}) can be rewritten as:
\begin{equation}
\label{vevs2}
\begin{array}{cccc}
\langle H \rangle=v,& \qquad
\langle H^\prime \rangle=v^\prime,
\end{array} \qquad
\langle \eta \rangle=v_{\eta} \left(\begin{array}{c}1\\1\end{array}\right)\,.
\end{equation}
The generator $B$ of $D_4$ acts as the identity on the singlet representations ($1_1$ for  $H$ and $1_3$ for $H^\prime$). On the other hand,
from equation (\ref{generators}) we also see that $B$, which is the generator of a $Z_2$ symmetry, leaves invariant the vector $\langle \eta \rangle$.   
In a compact $4\times 4$ form, such a generator can be written as: 
\begin{equation}
B_{4\times4}=\left(\begin{array}{cccc}1&0&0&0\\0&1&0&0\\0&0&0&1\\0&0&1&0\end{array}\right) \,,
\end{equation}
where it is understood that the first two entries work on the $H$ and $H^{'}$ fields and the others on the components of the $\eta$ field.
After performing the $3-4$ rotation that diagonalizes the corresponding entries in the mass matrices, the generator can be cast in a diagonal form:
\begin{equation}
\label{z2rem}
B_{4\times4}=\left(\begin{array}{cccc}1&0&0&0\\0&1&0&0\\0&0&1&0\\0&0&0&-1\end{array}\right) \,,
\end{equation}
showing the peculiar feature of the negative value in the $(4,4)$ entry.
The previous $2\times2$ rotation defines the mass eigenstates in the $D_4$ doublet subspace of eq.(\ref{etas}): 
\begin{eqnarray}
\eta_p&=&\frac{1}{\sqrt{2}}(\eta_2+\eta_1)\,,\label{etap}\nonumber \\ && \\ 
\eta_m&=&\frac{1}{\sqrt{2}}(\eta_2-\eta_1)\label{etam}\,,\nonumber
\end{eqnarray}
that, under the remnant $Z_2$ symmetry defined in eq.(\ref{z2rem}), transform as
\begin{equation}
\eta_p\rightarrow +\eta_p,\qquad \eta_m\rightarrow -\eta_m\,,
\end{equation}
with vevs\footnote{For a similar change of basis in $Q_6$ see \cite{Kaburaki:2010xc}.}
\begin{equation}
\langle\eta_m\rangle=0,\quad \langle\eta_p\rangle=\sqrt{2}v_{\eta}\,.
\end{equation}
{The same conclusions can be drawn working in a different basis where }
% Alternatively, assuming the generators for the two dimentional representation to be
\begin{equation}\label{generators2}
\begin{array}{cc}
A=\left(\begin{array}{cc}0&1\\-1&0\end{array}\right),&B=\left(\begin{array}{cc}1&0\\0&-1\end{array}\right)\,.
\end{array}
\end{equation} 
{In that case, $\eta$ develops a vev along the direction $(1,0)$ and the component that does
not take vev is charged under the remaining $Z_2$ parity}
% not take vev is charged under the remaining $Z_2$ patity }
% then in the doublet $\eta$ takes vev along the direction $(1,0)$ and the component that does
% not take vev is charged under the remaining $Z_2$ patity
\footnote{We thank Luis Lavoura to point out this possibility.}.

{Let us now make more transparent the role of the $Z_2^{f}$. If we  introduce linear and/or trilinear terms in the scalar potential 
transforming as  $1_2$ and $1_4$ representations (like $\phi$ in our case), the vev of $\eta$ would not be aligned along 
the direction $(1,1)$ and the residual $Z_2$ would be broken, causing the decay of the DM. In fact, one can add contractions of 
the form $(\eta^\dagger \eta)_{1_2}$ into $1_2$ singlet of $D_4$ and  the minimizing equations for the $\eta$ components would admit solutions only if $\vev{\eta_1}\ne \vev{\eta_2}$;
% For this reason we introduce 
{the auxiliarly symmetry
$Z_2^{f}$ (under which $\phi\to -\phi$) only allows quadratic and quartic terms in
$\phi$ which transform as a singlet $1_1$ and avoid the dangerous $Z_2$-breaking contractions. }

We see that $\eta_m$ is the only Higgs field charged under the $Z_2$ symmetry. This prevents any coupling with other Higgs fields and, 
considering also that the original $\eta$ field does not couple to quarks and charged lepton bilinears (but only to heavy right-handed neutrinos), 
the component of $\eta_m$ corresponding to the lightest $Z_2$-odd neutral spin zero
particle is the DM candidate of the model, namely the combination $\eta_1^\prime-\eta_2^\prime$. 
% Notice that the role of the $Z_2^{f}$ symmetry is that of eliminating from the scalar potential the coupling that would 
% prevent the alignment $\eta\sim(1,1)$.

Since the DM couplings with the standard model 
particles are very similar to the ones in ref. \cite{Hirsch:2010ru},  
we expect almost the same DM phenomenology as described in \cite{Boucenna:2011tj}, with $m_{DM}$ in the range $few~ GeV<m_{DM}<100~GeV$.

%Here we note that the term proportional to $\xi_3$ in the scalar potential gives, after that the scalar field $\phi$
%takes a vev, an additional mass term to $\mu_\eta$ that allows to have DM mass heavier, around $500$\,GeV as shown
%in ref\,\cite{Hirsch:2010ru}.

\section{Charged leptons and neutrinos}
\label{chargedsec} 
In the neutrino sector, the Dirac and Majorana mass matrices derived from eq.(\ref{lag}) read:
\begin{equation}
 m_{D}=v_{\eta}\,\left(\begin{array}{ccc}
y_1^\nu &y_1^\nu \\
y_2^\nu &-y_2^\nu \\
y_3^\nu &y_3^\nu 
\end{array}\right)\qquad
M_R=M_1\,\left(\begin{array}{ccc}
0&1\\
1&0
\end{array}\right)\,,
\end{equation}
and the resulting light neutrino mass matrix, from type-I seesaw mechanism $m_\nu=-m_D  M_R^{-1} m_D^T$, can be parametrized as:
\begin{equation}
m_\nu=
\left(
\begin{array}{ccc}
2 A^2 & 0&2 A C \\
0 & -2B^2& 0 \\
2 A C & 0& 2 C^2 \\
\end{array}
\right),
\end{equation}
where $A^2=(y_1^\nu v_{\eta})^2/M_1  $, $B^2= (y_2^\nu v_{\eta})^2/M_1 $ and  $C^2= (y_3^\nu v_{\eta})^2/M_1 $.
This matrix is diagonalized by: 
\begin{eqnarray}
U_\nu^T \cdot m_\nu \cdot U_\nu &=&D_\nu\,;
\end{eqnarray}
to find $U_\nu$, we first compute the eigenvalues $|m_{\nu_i}|$ and eigenvectors of 
$m_\nu$, related in the  following way:
\begin{eqnarray}
0 \to \left(
\begin{array}{c}
-\frac{|A| C^*}{A^*\,\sqrt{|A|^2+|C|^2}}\\
0\\
\frac{|A|}{\sqrt{|A|^2+|C|^2}} 
\end{array}
\right)\qquad 2|B|^2 \to\left(
\begin{array}{c}
0\\
1\\
0
\end{array}
\right)\qquad  2 (|A|^2+|C|^2) \to
\left(
\begin{array}{c}
\frac{A C^*}{C\,\sqrt{|A|^2+|C|^2}}\\
0\\
\frac{|C|}{\sqrt{|A|^2+|C|^2}} 
\end{array}
\right)\,.
\end{eqnarray}
The zero eigenvalue can be assigned to $m_{\nu_1}$ or $m_{\nu_3}$. In the latter case, for any  
ordering of the remaining mass eigenstates, the solar angle cannot be reproduced.
On the other hand, for $m_{\nu_1}=0$, we get an appropriate  $U_\nu$ if we associate
$|m_{\nu_2}|=2 (|A|^2+|C|^2)$ and $|m_{\nu_3}|=2  |B|^2$, with the condition 
$|B|^2 > (|A|^2+|C|^2)$ to fit the atmospheric mass difference. The resulting $U_\nu$
is then given by:
\begin{eqnarray}
\label{nurot}
U_{\nu}=
\left(
\begin{array}{ccc}
-\frac{|A| C^*}{A^*\,\sqrt{|A|^2+|C|^2}} &\frac{A C^*}{C\,\sqrt{|A|^2+|C|^2}} &0 \\
0& 0& 1 \\
\frac{|A|}{\sqrt{|A|^2+|C|^2}} & \frac{|C|}{\sqrt{|A|^2+|C|^2}} & 0 \\
\end{array}
\right) = \left(\begin{array}{ccc}
c_\odot & s_\odot &0 \\
0& 0& 1 \\
-s_\odot & c_\odot &0
\end{array}
\right)
\end{eqnarray}
with
\begin{equation}
\label{tans}
\tan \theta_{\odot}=\left|\frac{A}{C}\right|\,.
\end{equation}

Now we consider the charged sector.
Since we work in the left-right basis, the mass matrices  $M_\ell M_\ell^\dagger$ and $M^\dagger_\ell M_\ell$ are diagonalized by the unitary matrices 
$U_\ell$ and $V_\ell$, respectively:
\begin{eqnarray}
U_\ell^\dagger \cdot M_\ell M_\ell^\dagger \cdot U_\ell &=&D^2_\ell\\
V_\ell^\dagger \cdot M_\ell^\dagger M_\ell \cdot V_\ell &=&D^2_\ell\,,
\end{eqnarray}
where $D_\ell$ is the diagonal charged lepton mass matrix.

After electroweak symmetry breaking the mass matrix of the charged leptons has the form:
\begin{equation}
\label{charged}
M_{l}=\left(\begin{array}{ccc}
y_1^l v^\prime \varepsilon&0&0\\
0&y_2^lv\varepsilon&y_4^lv^\prime\varepsilon\\
0&y_5^lv^\prime&y_3^lv
\end{array}\right)
% =\left(\begin{array}{ccc}
% m_e&0 &0\\0 &a&b\\0&c&d
%\end{array}\right)
\end{equation} 
where we have defined $\langle \phi \rangle/\Lambda=\varepsilon$. The charged lepton masses are given by:
\begin{eqnarray}
 m_e &\approx& y_1^l v^\prime \varepsilon \nonumber \\
m_\mu &\approx& \varepsilon \frac{v^2  y_2^l y_3^l-v'^2  y_4^l y_5^l }{\sqrt{(v y_3^l)^2+ (v' y_5^l)^2}}  \\
m_\tau &\approx& \sqrt{(v y_3^l)^2+ (v' y_5^l)^2}\,.\nonumber
\end{eqnarray}
We can easily see that the muon mass is suppressed with respect to $m_\tau$ by a factor of $\varepsilon$ and 
enhanced  with respect to $m_e$ by roughly a factor of $v^\prime/v$, which is smaller than 1 (see below).
If we concentrate only on the $\mu-\tau$ submatrix, we can rewrite it as:
 \begin{equation}
M_{\mu\tau}=\left(\begin{array}{cc}
a&b\\c&d
\end{array}\right)\, ,\label{subm}
\end{equation}
where $a, b, c$ and $d$ are products of Yukawa couplings and vevs.
We can define two unitary rotations $V_L$ and $V_R$
\begin{equation}
V_L=\left(\begin{array}{cc}
\cos \theta_\ell& \sin  \theta_\ell\\
-\sin  \theta_\ell&\cos \theta_\ell
\end{array}\right),
\quad
V_R=\left(\begin{array}{cc}
\cos \theta_R& \sin  \theta_R\\
-\sin  \theta_R&\cos \theta_R
\end{array}\right)
\end{equation}
such that:
\begin{equation}
V_L^\dagger \, M_{\mu\tau}\, V_R=\left(\begin{array}{cc}
m_\mu^2&0\\0&m_\tau^2
\end{array}\right)\label{subm2}\,.
\end{equation}
Given the neutrino mixing matrix in eq.(\ref{nurot}), it is evident that the atmospheric angle originates from $V_L$ while the angle $\theta_R$ is a 
free parameter.
The global $3\times 3$ charged lepton mass matrix is then diagonalized by the following rotations:
\begin{equation}
U_\ell=
\left(\begin{array}{cc}
1&0\\
0&V_L
\end{array}\right),\qquad
V_\ell=
\left(\begin{array}{cc}
1&0\\
0&V_R
\end{array}\right).
\label{Uell}
\end{equation}
To get an estimate of the magnitude of the Yukawa parameters in (\ref{charged}) we proceed as follows.
The vevs $v$ and $v'$ are fixed from the minimization condition of the scalar potential, then the four Yukawa 
couplings $y^l_{2,3,4,5}$ (assumed to be real) are determined from the atmospheric angle, the $\mu$ and $\tau$
masses and the angle $\theta_R$. 
Let us assume $\theta_\ell = \pi/4$ (in good agreement with the experimental data); then, the conditions for having 
a $\mu\tau$ invariant submatrix (\ref{subm}) can be deduced using: 
\begin{equation}
M_{\mu\tau}M_{\mu\tau}^\dagger= V_L\left(\begin{array}{cc}
m_\mu^2&0\\0&m_\tau^2
\end{array}\right) V_L^\dagger\label{subm3}\,.
\end{equation}
We get:
\begin{eqnarray}
a^2+ b^2 &=& c^2+d^2 \nonumber \\
c^2+d^2&=&\frac{m_\tau^2+m_\mu^2}{2} \label{abc}\\
db+ca&=&\frac{m_\tau^2-m_\mu^2}{2}\nonumber\,.
\end{eqnarray}
In this way we can write three out of four  parameters (for instance, $a$, $b$ and $d$) in terms of the charged lepton masses,
the atmospheric angle (supposed to be maximal here) and $c$. The latter is related to the $\theta_R$ angle by:
\begin{eqnarray}
\tan 2\theta_R= -\frac{2 (a b+c d)}{a^2-b^2+c^2-d^2}\,.
\end{eqnarray}
It is easy to show that the system of equations in (\ref{abc}) has real solutions only of the form  
$a\sim c$ and $b\sim d$
from which we deduce \footnote{The relative hierarchy among the two groups of parameters depend on the choice of $\theta_R$.}:
\begin{equation}\label{yv}
y_2^lv\varepsilon\approx y_5^lv',\qquad y_3^lv\approx y_4^lv^\prime\varepsilon.
\end{equation}
% Then if $v'\ll v$ then $y_{2,3} \gg y_{4,5}$.
For $y_2^l\sim y_5^l$ the first relation implies $v^\prime/v\sim \varepsilon$ whereas the second one requires 
a moderate fine-tuning of order $y_3^l / y_4^l \sim \varepsilon^2$.

Finally, the lepton mixing matrix is given by:
\begin{equation}
U_{lep}=U_\ell^\dagger \cdot U_\nu=
\left(
\begin{array}{ccc}
c_\odot & s_\odot &0 \\
\sin \theta_\ell s_\odot & -c_\odot \sin \theta_\ell  & \cos \theta_\ell \\
-\cos \theta_\ell s_\odot & s_\odot \cos \theta_\ell  & \sin \theta_\ell 
\end{array}
\right)
\end{equation}
where, considering the relations in eq.(\ref{yv}), gives:
\begin{equation}
 \tan \theta_{23} \sim \frac{y_5^l}{y_3^l}\,\varepsilon\,.
\end{equation}
To reproduce the correct maximal mixing in the atmospheric sector we need $y_3^l/y_5^l\sim\varepsilon$.
We clearly see that our model predicts a vanishing $\theta_{13}$. 
%while the atmospheric angle is unpredicted and it remains a free parameter of the model. 
Since also $m_1^\nu=0$,  the effective mass entering the neutrinoless double beta decay assumes a particularly simple expression:
\begin{equation}
|m_{\beta\beta}|=|m_{\nu_2}| s_\odot^2=\sqrt{\Delta m^2_{\odot}} s_\odot^2,
\end{equation} 
where $\Delta m^2_{\odot}=|m_{\nu_2}|^2-|m_{\nu_1}|^2$, with numerical values in the interval:
\begin{equation}
0.00054 \,{\rm eV}\le |m_{\beta\beta}| \le 0.0012\, {\rm eV}\,.
\end{equation}

\section{Estimate of lepton flavor violating processes}
\label{fcnc}   
Since we have more than one $SU(2)$ Higgs doublet coupled to charged leptons, our model allows for lepton flavor violating processes (LFV) at 
tree level such as $\tau\rightarrow 3 \mu$ and $\tau\rightarrow \mu e e$ (see \cite{Kaneko:2006wi} for other examples of renormalizable 
models based on dihedral groups). Here we are not interested in  a full study of the LFV but just to show that the model 
prediction for them can be easily maintained below their upper bounds. 
% Below we give an estimation for some of LFV processes and verify that there exists a choice of the 
% model parameters which leaves them  in agreement with current experimental bounds. 
In the interaction basis, the Yukawa matrices $Y$ and $Y^\prime$ can be deduced from:
\begin{equation}\label{39} 
L\cdot Y\cdot l^c H+ L\cdot Y^\prime \cdot l^c H^\prime=L\left(\begin{array}{ccc}
0&0&0\\
0&y_2^l&0\\
0&0&y_3^l\end{array}\right)l^c H+L\left(\begin{array}{ccc}
y_1^l &0&0\\
0&0&y_4^l \\
0&y_5^l&0\end{array}\right)l^c H^\prime\,,\end{equation}
where we have reabsorbed $\varepsilon$ into $ y_1^l$, $ y_2^l$ and  $ y_4^l$. 
% In order to estimate the LFV processes we have to rotate the charged fermions to the mass basis. 
Once we rotate the fields $L$ and $l^c$ to the mass basis, the new Yukawa matrices $\tilde{Y}$ and $\tilde{Y^\prime}$ read:
\begin{equation}
\begin{array}{cc}
\tilde{Y}= U_\ell^\dagger \cdot Y \cdot V_\ell, & \;\tilde{Y^\prime}=U_\ell^\dagger\cdot Y^\prime \cdot V_\ell.
\end{array}
\end{equation}
They are not diagonal and contain non vanishing $\mu-\tau$ entries, as it can easily deduced using eq.(\ref{Uell}). 
The Higgs fields should also be expressed in the mass basis but we do not take this additional rotation into account 
since it would only introduce additional mixing angles as suppression factors in the branching fraction computations (we are then working in the case where the LFV processes are the largest allowed in our model).

As explained above, the Yukawa couplings $y_i$ are fixed from the value of fermion masses $m_e$, $m_\mu$ and 
$m_\tau$, the vevs $v$ and $v'$\footnote{In the computation of the branching ratios, we set $\sqrt{2}v_\eta=v^\prime$.} (determined by the potential parameters) and the arbitrary angle $\theta_R$ of the $V_\ell$ unitary matrix. 
To get realistic estimates, we performed a numerical simulation with the constraints defined below eq.(\ref{dmmass}).
It turns out that we can always find solutions with $ v>v^\prime$, which implies $\tilde Y^\prime> \tilde Y$, 
see eqs.(\ref{yv}) and (\ref{39}).
In this case, the decay width for the $\tau \rightarrow 3\mu$ process\,\footnote{The process $\tau\to \mu e e$
is suppressed by $\tilde{Y}'_{ee}$.}
 is approximated by:
\begin{equation}
\label{taubr}
\Gamma (\tau \rightarrow 3\mu )\approx
\frac{m_\tau^5\left(\tilde{Y}'_{\mu \mu}\tilde{Y}'_{\tau \mu}\right)^2}{6\times 2^9\pi^3m_{H^{\prime}}^4}\,.
\end{equation}
Numerical examples of the branching ratio $Br(\tau \rightarrow 3\mu )$ are given in Tab.\ref{br}, 
where we choose the vev $v$, the ratio $v/v^\prime$ and $m_{H^\prime}$ as independent variables. 
We fixed the value of $\theta_R$ to $\sin\theta_R = 0.9277$ for which the product of the Yukawas $\tilde{Y}'_{\mu \mu}\tilde{Y}'_{\tau \mu}$ is maximal 
and the branching ratios are the largest possible. 
\begin{table}[h!]
\begin{center}
\begin{tabular}{|c|c|c|c|c|l|}
\hline
$v$ (GeV)&$v/v^\prime$ &$m_{H^\prime}$ (GeV)&$m_{H}$ (GeV)&$m_{DM}$ (GeV)&$Br(\tau \rightarrow 3\mu )$\\
\hline
224  &3.6 &140&115&89&$ 8.1\times 10^{-9}$\\
\hline
225 &3.8&201&98&87&$ 2.5\times 10^{-9}$\\
\hline
225 &3.8&175&132&60&$3.1 \times 10^{-9}$\\
\hline
222 &3.5&206&118&84&$ 1.5\times 10^{-9}$\\
\hline
173 &1.5&266&223&75&$5.4 \times 10^{-11}$\\
\hline
\end{tabular}
\caption{\it \label{tab11}Branching ratio for the process $\tau \rightarrow 3\mu $ as deduced from our model. The experimental bound is 
$Br(\tau \rightarrow 3\mu )< 3.2\cdot 10^{-8}$ \cite{Nakamura:2010zzi}.}
\label{br}
\end{center}
\end{table}
As we can see, there is a region of the parameter space where the branching ratio for the tau decay is well 
below the experimental upper bound $Br(\tau \rightarrow 3\mu )< 3.2\cdot 10^{-8}$ \cite{Nakamura:2010zzi}. 
On the other hand, the first entry is very close to the upper limit, showing that a sector of the parameter space will be tested in the near 
future at the LHC.
% 
%  nevertheless, there is a region where the branching ratio is very close to the current experimental 
% upper bound. This means that the model could be tested in a near future, or at least will could give constraints on the parameter space, 
% which is related directly with the DM phenomenology. 

\section{The quark sector}
\label{quarks}
In this section we discuss the extension of our model to the quark sector.
The assignment of the quark fields to the irreducible representation of $D_4$ is listed in Tab.\ref{tab3}.
\begin{table}[h!]
\begin{center}
\begin{tabular}{|c|c|c|c|c|c|c|}
\hline
&$Q_1$&$Q_2$&$Q_3$&$q_1^c$&$q_2^c$&$q_3^c$\\
\hline
$SU(2)$&2&2&2&1&1&1\\
\hline
$D_4$ &$1_4$ &$1_2$&$1_1$&$1_4$&$1_2$&$1_1$\\
\hline
$Z_2^f$ &$+$ &$-$&$+$&$+$&$-$&$+$\\
\hline
\end{tabular}
\caption{\it Quark assignments in our model.}\label{tab3}
\end{center}
\end{table}

The Lagrangian for the down-type quarks reads as follows:
\begin{equation}\label{lagdown}
\begin{array}{lcl}
\mathcal{L}_{down}&=&y^d_{1} \,{Q}_1 q_{_1}^c H +
y^d_{2} \,{Q}_1 q_{_2}^c H' + y^d_{3}\, {Q}_2 q_{_1}^c H'+
y^d_{4} \,{Q}_2 q_{_2}^c H + y^d_{5}\, {Q}_3 q_{_3}^c H+\\
&+&\frac{\phi}{\Lambda}\left(y^d_{6}{Q}_1 q_{_3}^c H'+y^d_{7}{Q}_2 q_{_3}^c H+y^d_{8}{Q}_3 q_{_1}^c H'
+y^d_{9}{Q}_3 q_{_2}^c H\right).\end{array}
\end{equation}
For the up-type quarks the Lagrangian has the same structure with the obvious replacements $y^d_{i}\to y^u_{i} $
and $(H,H') \to(\tilde H,\tilde H')$, where $\tilde H= -i \tau_2\,H^\dagger$. The mass matrices are then:
\begin{equation}
m_{u,d}=
\left(
\begin{array}{ccc}
y_1^{u,d} v  & y_2^{u,d} v^\prime &  y_6^{u,d}v^\prime\varepsilon\\
y_3^{u,d}  v^\prime &y_4^{u,d} v & y_7^{u,d}v \varepsilon \\
y_8^{u,d}v^\prime\varepsilon &y_9^{u,d}v\varepsilon & y_5^{u,d} v \\
\end{array}
\right)\,.
\end{equation}
With such a texture we can easily fit the quark masses and the CKM mixing angles.
In particular, given that $v>v^\prime$ (as discussed before eq.(\ref{taubr})), we can fix $y_{1,4,5}^{u,d}$ 
in such a way that $m_1^{u,d}\sim y_1^{u,d} v$,
$m_2^{u,d}\sim y_4^{u,d} v$ and $m_3^{u,d}\sim y_5^{u,d} v$. 
Then the Cabibbo angle is given by:
\begin{equation}
\theta_C\sim \left( \frac{y_2^u}{y_4^u}- \frac{y_2^d}{y_4^d}\right)\frac{v'}{v}\,,
\end{equation}
and it can be fit to its experimental value for a suitable choice of the vev ratio $v'/v$ 
(with $y_2^{u,d}$ of the same order of $y_4^{u,d}$). Taking $y_{6,7}^{u,d}$ of the same order of $y_5^{u,d}$,
we also have $V_{ub}\approx \varepsilon v'/v $  and  $V_{cb}\approx \varepsilon $  (which fixes $\varepsilon\sim \cal O$(0.04)).
\\
%
%The Cabibbo angle can be obtained in general from a cancellation between the  down and up quark sectors, but we can have limit
%where it arises only from one of the two sectors.
%
Since the Cabibbo mixing arises from both the up and down sectors, we can have $s-d$ and 
$c-u$ tree-level transitions mediated by Higgses. This implies 
that decays like $K^{+,0}\to \pi^{+,0}l\bar{l}$ (in the down sector) or 
$D^{+,0}\to \pi^{+,0}l\bar{l}$ and $D^{+}_s\to K^{+}l\bar{l}$ (in the up sector) can exceed their experimental bounds.
Since the coupling $\tilde{Y}'_{ee}$ is suppressed by the electron mass, the pairs $l\bar{l}$ can be
$\mu^-\mu^+$ whereas the case $\mu^\pm\tau^\mp$ is kinematically excluded. 
The tree-level transitions $b-s$, $b-d$ are suppressed by $\varepsilon$ and we only consider  
$B^+\to K^+\mu^+\mu^-$ and $B^+\to \pi^+\mu^+\mu^-$.
In order to give an estimate of such processes, we work in the worst case of unity 
mixings and (adimensional, that is stripped of the meson masses) form factors. We then have:
\begin{equation}\label{meson} 
\begin{array}{lll}
\Gamma(K^{+,0}\to \pi^{+,0}\mu^+ \mu^-)&\approx & 
\frac{1}{3072 \pi^3}\left( \frac{m_K^5}{m_{H'}^4}\right)|\tilde{Y}'_{ds}\tilde{Y}'_{\mu\mu}|^2,\\
&&\\
\Gamma(D^{+,0}\to \pi^{+,0}\mu^+ \mu^-)&\approx & 
\frac{1}{3072 \pi^3}\left( \frac{m_D^5}{m_{H'}^4}\right)|\tilde{Y}'_{cu}\tilde{Y}'_{\mu\mu}|^2,\\
&&\\
\Gamma(D^{+}_s\to K^+ \mu^+ \mu^-)&\approx & 
\frac{1}{3072 \pi^3}\left( \frac{m_{D_s}^5}{m_{H'}^4}\right)|\tilde{Y}'_{cu}\tilde{Y}'_{\mu\mu}|^2.\\
&&\\
\Gamma(B^{+}\to K^+ \mu^+ \mu^-)&\approx & 
\frac{1}{3072 \pi^3}\left( \frac{m_{B}^5}{m_{H'}^4}\right)|\tilde{Y}'_{bs}\tilde{Y}'_{\mu\mu}|^2.\\
&&\\
\Gamma(B^{+}\to \pi^+ \mu^+ \mu^-)&\approx & 
\frac{1}{3072 \pi^3}\left( \frac{m_{B}^5}{m_{H'}^4}\right)|\tilde{Y}'_{bd}\tilde{Y}'_{\mu\mu}|^2.
\end{array}
\end{equation}
We observe that for $v\sim 224\,$GeV  (the worst point in Tab.\ref{tab11}), 
the couplings $y^l_{2,3,4,5}$ in eq.(\ref{39}) should be of ${\cal O}$($10^{-2}$) to reproduce the $\tau$ mass and
then $\tilde{Y}_{\mu\mu}\sim 10^{-2}$. 
In the down sector we have $y_5^d\sim 2\cdot 10^{-2}$, $y_4^d\sim 6\cdot 10^{-4}$, $y_1^d\sim 3\cdot 10^{-5}$
and $y_2^d\sim y_4^d$ and therefore $\tilde{Y}'_{ds}\sim 6\cdot 10^{-4}$. A similar reasoning in the up sector gives
$\tilde{Y}'_{cu}\sim 7\cdot 10^{-3}$. Using the above values, we computed the branching ratios for the meson decay processes in 
eq.(\ref{meson}). They are summarized in Tab.\ref{tab4}.
\begin{table}[h!]
\begin{tabular}{|l|c|c|}
\hline
$decay$ & model prediction & experimental bounds \cite{Nakamura:2010zzi}\\
\hline
$\text{Br}(K^{+}\to \pi^{+}\mu^+ \mu^-)$ &$5.5\cdot 10^{-9}$ &$<8.1 \cdot 10^{-8}$ \\
$\text{Br}(K^{0}\to \pi^{0}\mu^+ \mu^-)$ &$1.2\cdot 10^{-9}$ &$<2.9 \cdot 10^{-9}$ \\
$\text{Br}(D^{+}\to \pi^{+}\mu^+ \mu^-)$ &$1.3\cdot 10^{-8}$  &$<3.9 \cdot 10^{-6}$ \\
$\text{Br}(D^{0}\to \pi^{0}\mu^+ \mu^-)$ &$5.2\cdot 10^{-9}$ &$<1.8 \cdot 10^{-4}$ \\
$\text{Br}(D^{+}_s\to K^{+}\mu^+ \mu^-)$ &$8.3\cdot 10^{-9}$ &$<3.6 \cdot 10^{-5}$ \\
$\text{Br}(B^{+}\to K^{+}\mu^+ \mu^-)$   &$6.4\cdot 10^{-8}$ &$<5.2 \cdot 10^{-7}$ \\
$\text{Br}(B^{+}\to \pi^{+}\mu^+ \mu^-)$ &$3.2\cdot 10^{-10}$ &$<1.4 \cdot 10^{-6}$ \\
\hline
\end{tabular}
\caption{\it Branching fraction estimates for some interesting processes in our model. We fixed $m_{H'}=140$ GeV,
$v=224$ and $\tilde{Y}'_{\mu\mu}=0.0145$.}\label{tab4}
\end{table} 

We clearly see that all branching ratios are below their upper limits. We have also checked that the mass difference 
in the kaon system, driven by the $K^0-\bar K^0$ oscillation, is around $10^{-14}$ GeV, to be compared with the experimental value
$\sim 10^{-12}$ GeV.
We stress that, even if we have taken a particular point in the parameter space to make our estimates, the exercise can be 
repeated for different input values with similar conclusions.

\section{Conclusions}
\label{conclusions}
In this paper we have discussed an extension of the standard model based on the non-abelian discrete group $D_4$. 
We introduced three more $SU(2)$ Higgs doublets, a combination of them giving a good dark matter candidate,  one standard model singlet 
and only two right-handed neutrinos, a remarkable feature if compared 
with the models in \cite{Hirsch:2010ru,Meloni:2010sk,Boucenna:2011tj}. The stability of the DM candidate 
is not imposed ad-hoc but directly follows from 
the remnant $Z_2$ subgroup of the broken $D_4$, as we explained in details. Within the same framework, we incorporated a description of the charged leptons 
and neutrinos, showing that the normal hierarchy (with $m_{\nu_1}=0$) and a vanishing $\theta_{13}$ are natural predictions of our model. 
This also allows to get a range of values for the effective mass $m_{\beta\beta}$, which turns out to be in the interval $\left[0.5,1.2\right]\,\cdot 10^{-3}$ eV.
On the other hand, 
the solar and atmospheric angles are free parameters that can be easily fixed to their corresponding experimental values. 
We have carefully checked that, in a large portion of the parameter space, the model does not conflict with the upper bounds on some lepton flavor violating
processes, like $\tau \rightarrow 3\mu$. Finally, we extended the $D_4$ symmetry to the quark sector,
showing that {the correct order of magnitude for the CKM angles} can be easily reproduced assigning the quark fields to non trivial 
representation of $D_4$. Three-level flavor changing neutral current processes, generated by non-vanishing 
{off diagonal Yukawa couplings}, can be maintained below their current experimental bounds.

\section{Acknowledgments}
We are grateful to Luis Lavoura for useful comments.
Work of E.P. and S.M.  was supported by the Spanish MICINN under grants
FPA2008-00319/FPA and MULTIDARK Consolider CSD2009-00064, by
Prometeo/2009/091, by the EU grant UNILHC PITN-GA-2009-237920.
S.M. was supported by a Juan de la Cierva contract. E.P. was supported by CONACyT.
D.M. acknowledges MIUR (Italy) for partial financial support under the contract PRIN08.


\begin{thebibliography}{9}
\bibitem{Bertone:2004pz}
  G.~Bertone, D.~Hooper and J.~Silk,
  %``Particle dark matter: Evidence, candidates and constraints,''
  Phys.\ Rept.\  {\bf 405} (2005) 279
  [arXiv:hep-ph/0404175].
  %%CITATION = PRPLC,405,279;%%
\bibitem{Bertonebook}
  Bertone, G. (ed.) 
%  ``Dark Matter: Observations, Models and Searches''
``Particle dark matter: Evidence, candidates and constraints'',
(Cambridge Univ. Press, 2010)
  %%CITATION = PRPLC,405,279;%%

\bibitem{Taoso:2007qk}
  M.~Taoso, G.~Bertone, A.~Masiero,
  %``Dark Matter Candidates: A Ten-Point Test,''
  JCAP {\bf 0803}, 022 (2008).
  [arXiv:0711.4996 [astro-ph]].


%\cite{Hambye:2010zb}
\bibitem{Hambye:2010zb}
  T.~Hambye,
  %``On the stability of particle dark matter,''
  arXiv:1012.4587 [hep-ph].
  %%CITATION = ARXIV:1012.4587;%%



\bibitem{Hirsch:2010ru}
  M.~Hirsch, S.~Morisi, E.~Peinado and J.~W.~F.~Valle,
  %``Discrete dark matter,''
  Phys.\ Rev.\  D {\bf 82}, 116003 (2010)
  [arXiv:1007.0871 [hep-ph]].
  %%CITATION = PHRVA,D82,116003;%%
%\cite{Meloni:2010sk}
\bibitem{Meloni:2010sk}
  D.~Meloni, S.~Morisi and E.~Peinado,
  %``Neutrino phenomenology and stable dark matter with A4,''
  Phys.\ Lett.\  B {\bf 697} (2011) 339
  [arXiv:1011.1371 [hep-ph]].
  %%CITATION = PHLTA,B697,339;%%
%\cite{Boucenna:2011tj}
\bibitem{Boucenna:2011tj}
  M.~S.~Boucenna, M.~Hirsch, S.~Morisi, E.~Peinado, M.~Taoso and J.~W.~F.~Valle,
  %``Phenomenology of Dark Matter from A4 Flavor Symmetry,''
  arXiv:1101.2874 [hep-ph].
  %%CITATION = ARXIV:1101.2874;%%







%\cite{Haba:2010ag}
\bibitem{Haba:2010ag}
  N.~Haba, Y.~Kajiyama, S.~Matsumoto, H.~Okada and K.~Yoshioka,
  %``Universally Leptophilic Dark Matter From Non-Abelian Discrete Symmetry,''
  Phys.\ Lett.\  B {\bf 695}, 476 (2011)
  [arXiv:1008.4777 [hep-ph]].
  %%CITATION = PHLTA,B695,476;%%


%\cite{Kajiyama:2010sb}
\bibitem{Kajiyama:2010sb}
  Y.~Kajiyama and H.~Okada,
  %``T(13) Flavor Symmetry and Decaying Dark Matter,''
  arXiv:1011.5753 [hep-ph].
  %%CITATION = ARXIV:1011.5753;%%


%\cite{Daikoku:2010ew}
\bibitem{Daikoku:2010ew}
  Y.~Daikoku, H.~Okada and T.~Toma,
  %``Seesaw Mechanism Confronting PAMELA in S_4 Flavor Symmetric Extra U(1)
  %Model,''
  arXiv:1010.4963 [hep-ph].
  %%CITATION = ARXIV:1010.4963;%%
%\cite{Adulpravitchai:2011ei}
\bibitem{Adulpravitchai:2011ei}
  A.~Adulpravitchai, B.~Batell and J.~Pradler,
  %``Non-Abelian Discrete Dark Matter,''
  arXiv:1103.3053 [hep-ph].
  %%CITATION = ARXIV:1103.3053;%%


%\cite{LopezHonorez:2006gr}
\bibitem{LopezHonorez:2006gr}
  L.~Lopez Honorez, E.~Nezri, J.~F.~Oliver and M.~H.~G.~Tytgat,
  %``The inert doublet model: An archetype for dark matter,''
  JCAP {\bf 0702}, 028 (2007)
  [arXiv:hep-ph/0612275].
  %%CITATION = JCAPA,0702,028;%%





%\bibitem{Frigerio:2009wf}
%  M.~Frigerio, T.~Hambye,
  %``Dark matter stability and unification without supersymmetry,''
%  Phys.\ Rev.\  {\bf D81}, 075002 (2010).
%  [arXiv:0912.1545 [hep-ph]].


%\cite{Gu:2010ys}
%\bibitem{Gu:2010ys}
%  P.~H.~Gu, E.~Ma and U.~Sarkar,
  %``Pseudo-Majoron as Dark Matter,''
%  arXiv:1004.1919 [hep-ph].
  %%CITATION = ARXIV:1004.1919;%%


%\cite{Blum:2007nt}
\bibitem{Blum:2007nt}
%\cite{Grimus:2003kq}
%\bibitem{Grimus:2003kq}
  W.~Grimus and L.~Lavoura,
  %``A Discrete symmetry group for maximal atmospheric neutrino mixing,''
  Phys.\ Lett.\  B {\bf 572}, 189 (2003)
  [arXiv:hep-ph/0305046];
  %%CITATION = PHLTA,B572,189;%%
%\bibitem{Grimus:2004rj}
  W.~Grimus, A.~S.~Joshipura, S.~Kaneko, L.~Lavoura and M.~Tanimoto,
  %``Lepton mixing angle theta(13) = 0 with a horizontal symmetry D(4),''
  JHEP {\bf 0407} (2004) 078
  [arXiv:hep-ph/0407112];
  %%CITATION = JHEPA,0407,078;%%
%\cite{Babu:2004tn}
%\bibitem{Babu:2004tn}
  K.~S.~Babu and J.~Kubo,
  %``Dihedral families of quarks, leptons and Higgses,''
  Phys.\ Rev.\  D {\bf 71}, 056006 (2005)
  [arXiv:hep-ph/0411226].
  %%CITATION = PHRVA,D71,056006;%%
%\cite{Ishimori:2008ns}
%\cite{Ko:2007dz}
%\bibitem{Ko:2007dz}
  P.~Ko, T.~Kobayashi, J.~h.~Park and S.~Raby,
  %``String-derived D(4) flavor symmetry and phenomenological implications,''
  Phys.\ Rev.\  D {\bf 76}, 035005 (2007)
  [Erratum-ibid.\  D {\bf 76}, 059901 (2007)]
  [arXiv:0704.2807 [hep-ph]];
  %%CITATION = PHRVA,D76,035005;%%
  A.~Blum, C.~Hagedorn and A.~Hohenegger,
  %``theta_C from the Dihedral Flavor Symmetries D_7 and D_14,''
  JHEP {\bf 0803}, 070 (2008)
  [arXiv:0710.5061 [hep-ph]];
  %%CITATION = JHEPA,0803,070;%%
%\cite{Blum:2007jz}
%\bibitem{Blum:2007jz}
  A.~Blum, C.~Hagedorn and M.~Lindner,
  %``Fermion Masses and Mixings from Dihedral Flavor Symmetries with Preserved
  %Subgroups,''
  Phys.\ Rev.\  D {\bf 77}, 076004 (2008)
  [arXiv:0709.3450 [hep-ph]];
  %%CITATION = PHRVA,D77,076004;%%
%\cite{Ishimori:2008gp}
%\bibitem{Ishimori:2008gp}
  H.~Ishimori, T.~Kobayashi, H.~Ohki, Y.~Omura, R.~Takahashi and M.~Tanimoto,
  %``D(4) Flavor Symmetry for Neutrino Masses and Mixing,''
  Phys.\ Lett.\  B {\bf 662}, 178 (2008)
  [arXiv:0802.2310 [hep-ph]];
  %%CITATION = PHLTA,B662,178;%%
%\bibitem{Ishimori:2008ns}
  H.~Ishimori, T.~Kobayashi, H.~Ohki, Y.~Omura, R.~Takahashi and M.~Tanimoto,
  %``Soft supersymmetry breaking terms from D(4) x Z(2) lepton flavor
  %symmetry,''
  Phys.\ Rev.\  D {\bf 77}, 115005 (2008)
  [arXiv:0803.0796 [hep-ph]];
  %%CITATION = PHRVA,D77,115005;%%

 A.~Adulpravitchai, A.~Blum and C.~Hagedorn,
  %``A Supersymmetric D4 Model for mu-tau Symmetry,''
  JHEP {\bf 0903}, 046 (2009)
  [arXiv:0812.3799 [hep-ph]].
  %%CITATION = JHEPA,0903,046;%%
%\cite{Hagedorn:2010mq}
%\bibitem{Hagedorn:2010mq}
  C.~Hagedorn, R.~Ziegler,
  %``$\mu-\tau$ Symmetry and Charged Lepton Mass Hierarchy in a Supersymmetric $D_4$ Model,''
  Phys.\ Rev.\  {\bf D82}, 053011 (2010).
  [arXiv:1007.1888 [hep-ph]].




%\cite{Ishimori:2010au}
\bibitem{Ishimori:2010au}
  H.~Ishimori, T.~Kobayashi, H.~Ohki, Y.~Shimizu, H.~Okada and M.~Tanimoto,
  %``Non-Abelian Discrete Symmetries in Particle Physics,''
  Prog.\ Theor.\ Phys.\ Suppl.\  {\bf 183}, 1 (2010)
  [arXiv:1003.3552 [hep-th]].
  %%CITATION = PTPSA,183,1;%%



\bibitem{Grimus:2008nb}
  W.~Grimus, L.~Lavoura, O.~M.~Ogreid, P.~Osland,
  %``The Oblique parameters in multi-Higgs-doublet models,''
  Nucl.\ Phys.\  {\bf B801}, 81-96 (2008).
  [arXiv:0802.4353 [hep-ph]].
%\cite{Kaburaki:2010xc}
\bibitem{Kaburaki:2010xc}
  Y.~Kaburaki, K.~Konya, J.~Kubo and A.~Lenz,
  %``Triangle Relation of Dark Matter, EDM and CP Violation in B0 Mixing in a
  %Supersymmetric Q6 Model,''
  arXiv:1012.2435 [hep-ph].
  %%CITATION = ARXIV:1012.2435;%%


%%%%%%%%FCNC

%\cite{Kaneko:2006wi}
\bibitem{Kaneko:2006wi}
  S.~Kaneko, H.~Sawanaka, T.~Shingai, M.~Tanimoto and K.~Yoshioka,
  %``Flavor Symmetry and Vacuum Aligned Mass Textures,''
  Prog.\ Theor.\ Phys.\  {\bf 117} (2007) 161
  [arXiv:hep-ph/0609220];
  %%CITATION = PTPKA,117,161;%%
%\cite{Mondragon:2007af}
%\bibitem{Mondragon:2007af}
  A.~Mondragon, M.~Mondragon and E.~Peinado,
  %``Lepton masses, mixings and FCNC in a minimal S(3)-invariant extension of
  %the Standard Model,''
  Phys.\ Rev.\  D {\bf 76} (2007) 076003
  [arXiv:0706.0354 [hep-ph]];
  %%CITATION = PHRVA,D76,076003;%%
%\cite{Kifune:2007fj}
%\bibitem{Kifune:2007fj}
  N.~Kifune, J.~Kubo and A.~Lenz,
  %``Flavor Changing Neutral Higgs Bosons in a Supersymmetric Extension based on
  %a $Q_{6}$ Family Symmetry,''
  Phys.\ Rev.\  D {\bf 77} (2008) 076010
  [arXiv:0712.0503 [hep-ph]].
  %%CITATION = PHRVA,D77,076010;%%


\bibitem{Nakamura:2010zzi}
  K.~Nakamura {\it et al.} [ Particle Data Group Collaboration ],
  %``Review of particle physics,''
  J.\ Phys.\ G {\bf G37}, 075021 (2010).


%\bibitem{fede}
%F.~Bazzocchi, private comunication.





\end{thebibliography}
\end{document}